\documentstyle[epsfig]{aipproc}

\def \deg      {$^o$}
\def \gray     {$\gamma$-ray}

\def \sig      {$\sigma$}

\begin{document}
\title{COMPTEL Time-Averaged All-Sky Point Source Analysis}

\author{W. Collmar$^*$, V. Sch\"onfelder$^*$, A.W. Strong$^*$, H. Bloemen$^+$,\\
        W. Hermsen$^+$, M. McConnell$^{\dagger}$, J. Ryan$^{\dagger}$,
        K. Bennett$^@$
       }
\address{
$^*$Max-Planck-Institut f\"ur extraterrestrische Physik, Postfach 1603, 85740 Garching, Germany\\
$^+$SRON-Utrecht, Sorbonnelaan 2, NL-3584 CA Utrecht, The Netherlands\\
$^{\dagger}$Universtity of New Hampshire, Durham NH 03824-3525,USA\\
$^@$Astrophysics Division, ESTEC, NL-2200 AG Noordwijk, The Netherlands\\
  }

\maketitle

\vspace{-0.3cm}
\begin{abstract}
We use all COMPTEL data from the beginning of the CGRO mission (April '91) 
up to the end of CGRO Cycle 6 (November '97) to carry out all-sky point
source analyses in the four standard COMPTEL energy bands for different 
time periods. We apply our standard maximum-likelihood method to generate all-sky significance and flux maps for point sources by subtracting off the diffuse emission components via model fitting. In addition, fluxes of known sources have been determined for individual CGRO Phases/Cycles to generate lightcurves with a time resolution of the order of one year. The goal of the analysis is to derive quantitative results -- significances, fluxes,
light curves -- of our brightest and most significant sources such as 3C 273, and to search for additional new COMPTEL sources, showing up in time-averaged maps only.
\end{abstract}

\section*{Introduction}
The imaging COMPTEL experiment aboard CGRO is the pioneering satellite 
experiment of the MeV-sky ($\sim$1 - 30~MeV). For a detailed description of COMPTEL see \cite{Schonfelder93}. One of COMPTEL's prime 
goals is the generation of all-sky maps, which provide a summary 
on the MeV-sky in total. This goal has been achieved by e.g. \cite{Strong97}, \cite{Bloemen99} who generated maximum-entropy all-sky
images and by \cite{Blom97}, who generated the first COMPTEL all-sky maximum-likelihood maps, which -- compared to maximum-entropy ones -- 
have the advantage of providing quantitative results like significances and fluxes of source features. 
Here we present all-sky maximum-likelihood maps from which models of the 
diffuse emission have been removed. Our emphasis is on AGN. For a discussion 
on the method see \cite{Bloemen00} in these proceedings. 

The main analysis goals are
1) to derive a summary of known COMPTEL point sources,
2) to search for further point sources, 
3) to derive time-averaged quantitative parameters ('first order') of our
 brightest point sources, i.e., significances, fluxes, MeV-spectra,
 and possible time variability, 
 and 4) to further investigate our data and analysis methods.

\section*{Data and Analysis Method}
\begin{figure}[p!]
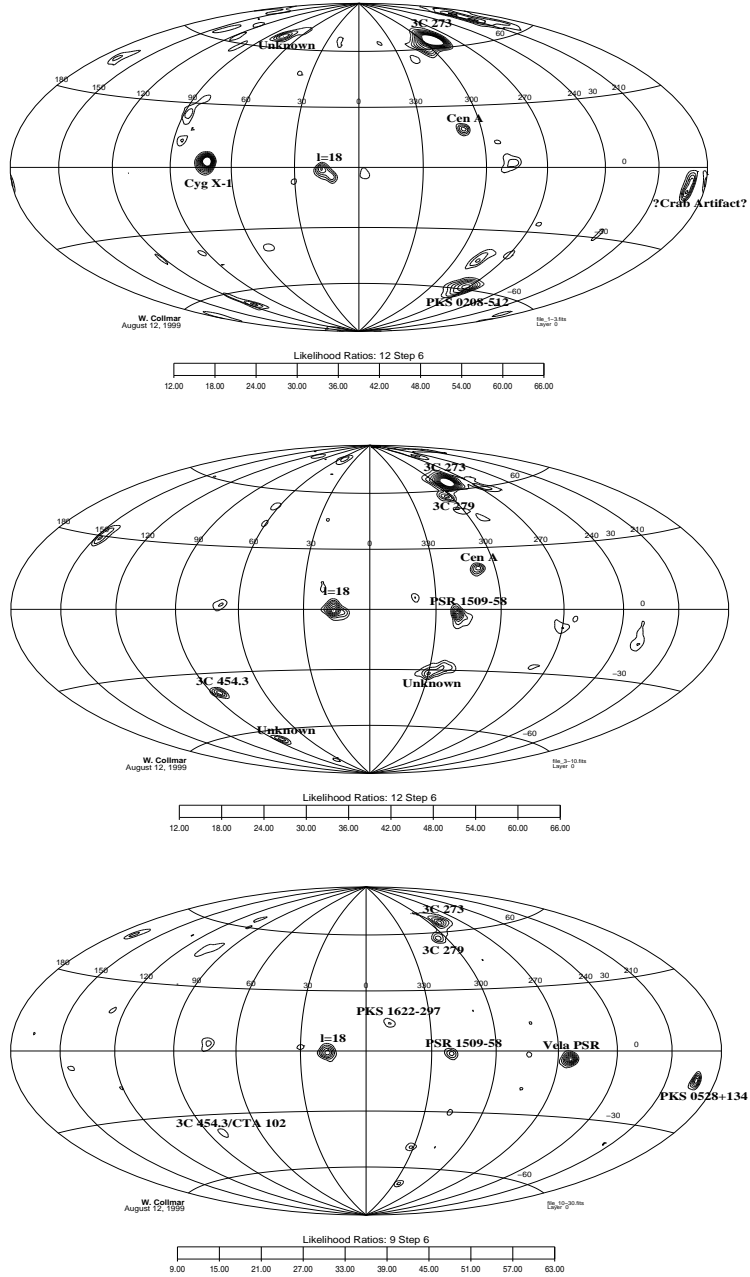
 
\centerline{\epsfig{file=skymap_xfig_1-3.eps_bw,height=2.3in,width=4.0in}}
\centerline{\epsfig{file=skymap_xfig_3-10.eps_bw,height=2.3in,width=4.0in}}
\centerline{\epsfig{file=skymap_xfig_10-30.eps_bw,height=2.3in,width=4.0in}}
\vspace{10pt}
\caption{
COMPTEL time-averaged maximum-likelihood point source all-sky maps in the 1-3 (upper), 3-10 (middle), and 10-30~MeV (lower panel) energy bands for the
time period April '91 to November '97, i.e. CGRO phases I-VI.
The galactic and extragalactic diffuse emission as well as the emission
from the Crab have been subtracted off via model fitting.
For the 1-3~MeV and the 3-10~MeV significance maps 
the contour lines start at a likelihood ratio value of 12 ($\sim$3.5\sig\ for
a known source; $\chi^{2}_{1}$-statistics) and for the 10-30~MeV map at a likelihood ratio value of 9 (3.0\sig\ for a known source) with steps of 6 for all maps. The most significant source features are labeled.  
}
\label{fig1}
\end{figure}

\begin{figure}[t!] 
\epsfig{file=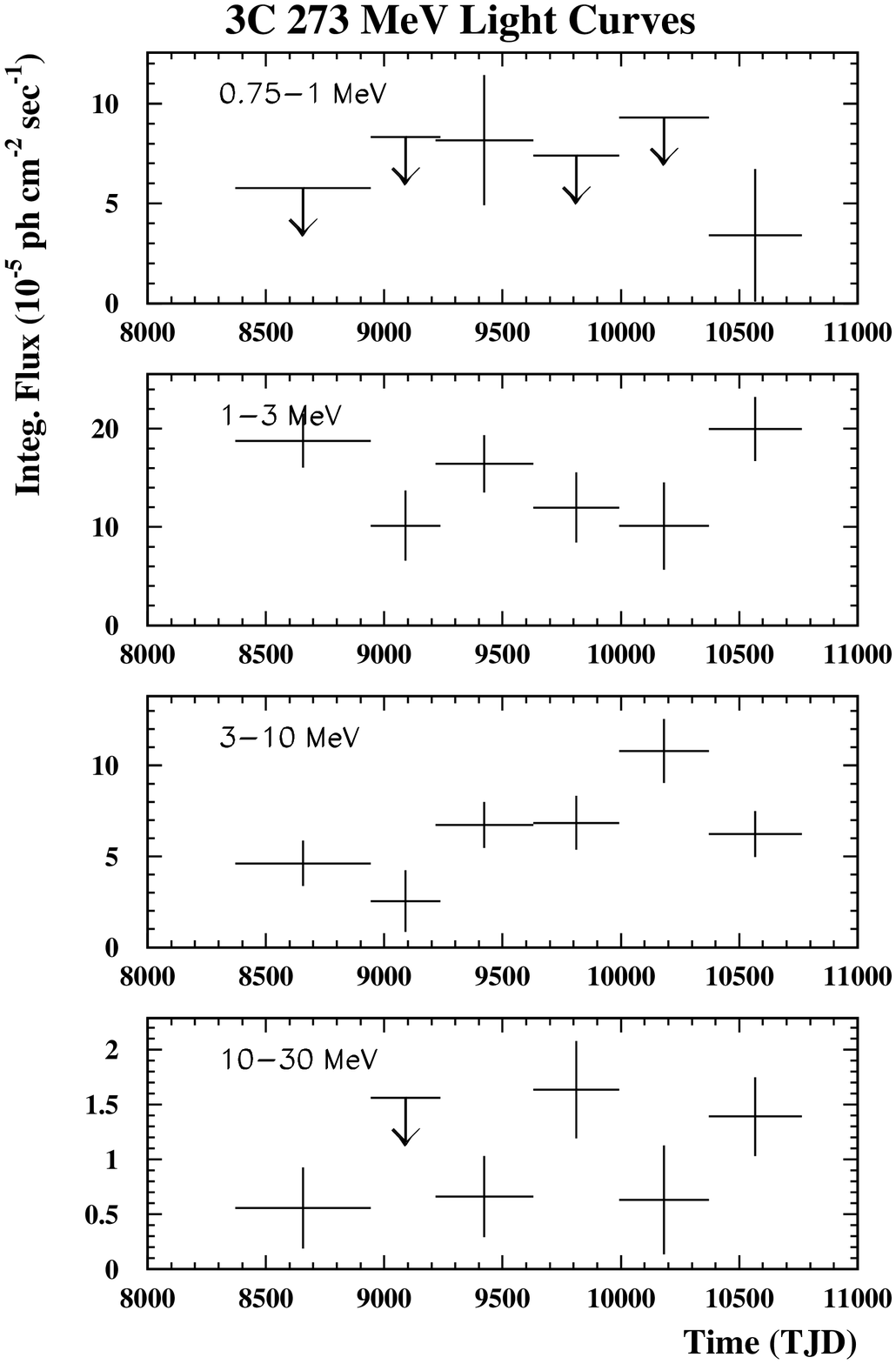,height=4.0in,width=2.5in}
\hfill
\epsfig{file=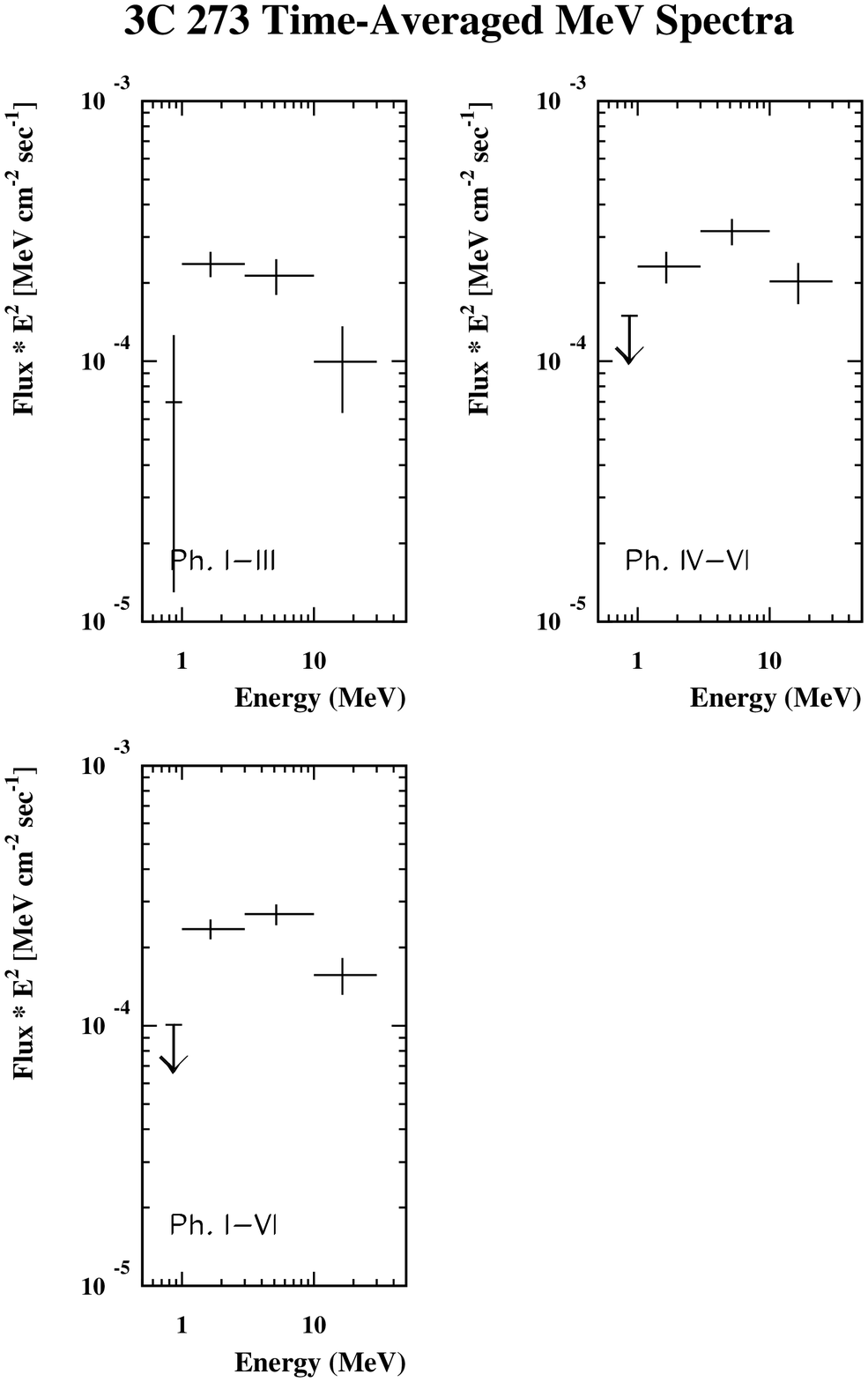,height=4.0in,width=2.5in}
\vspace{10pt}
\caption{
Left: COMPTEL light curves of the quasar 3C~273 in the 4 standard bands as derived from
all-sky flux fitting. Each flux point is averaged over an individual CGRO phase.
The error bars are 1\sig\ and the upper limits are 2\sig. 
\newline
Right: COMPTEL MeV spectra of 3C~273 for the sum of all CGRO phase I-III (April '91 - Oct. '94), IV-VI (Oct. '94 - Nov. '97), and I-VI (April '91 - Nov. '97) data  
as derived from all-sky flux fitting. The error bars are 1\sig\ 
and the upper limits are 2\sig.
}
\label{fig2}
\end{figure}

Using all data from the beginning of the CGRO mission (April '91) to the 
end of CGRO Cycle VI (Nov. '97), we generated a consistent database
of relevant COMPTEL data sets (events, exposure, geometry) for
individual CGRO viewing periods (VPs) in the 4 standard energy
bands (0.75-1, 1-3, 3-10, 10-30~MeV) in galactic coordinates by applying consistent data selections. 
This database was supplemented by relevant data sets containing models describing the galactic diffuse \gray\ emission (HI, CO, and inverse-Compton components) and the isotropic extragalactic \gray\ background emission.
To check for time variability of \gray\ sources these data sets were combined 
for different time periods: the six individual CGRO Phases/Cycles, the sum 
of all data (CGRO Phases I-VI; April '91 - Nov. '97) as well as the first 
(CGRO Phases I-III; April '91 - Oct. '94) and the second half (CGRO Phases IV-VI; Oct. '94 - Nov. '97). 
Each set of all-sky data is analysed by our standard maximum-likelihood 
method which simultaneously 'handles' individual VPs, generates, iteratively, a background model (see \cite{Bloemen94}), and finally generates significance
and flux maps and/or significances and fluxes for individual sources.      
Because we are interested in point sources, the diffuse emission is always 
removed in the fitting procedure (e.g. Figure~\ref{fig1}).  
For the derivation of the source fluxes (see Figure~\ref{fig2} as an example), 
the point sources of interest (e.g. 3C~273, Cyg~X-1) have additionally been included in the fitting procedure.    
We like to mention however, that the results 
derived by such all-sky fits should be considered correct to 
first order only. To derive final/optimal results for a particular source, 
a dedicated analysis has to be carried out, which e.g. makes several 
cross checks by applying different background models and would take into 
account the presence of other source features in the region of interest. Also, along the galactic plane the results depend on the 'goodness'
of the applied diffuse emission models for the MeV-band.

\section*{Results}

The significance maps in Figure~\ref{fig1}, which contain all data of the first 6.5 years of the COMPTEL mission, are the first COMPTEL all-sky point source 
maps in the continuum bands. They provide a summary of the on-average brightest and most significant MeV-sources. Similar maps focussing on the Galactic plane
only are given elsewhere in these proceedings ([5]).  
The Crab -- for display reasons removed in all maps of Figure~\ref{fig1} --
is by far the most significant COMPTEL point source. In the 1-3~MeV 
band for example it reaches a significance of $\sim$110\sig\ (i.e. a likelihood ratio of $\sim$12000) for the CGRO Phase I-VI period.
With significances of $\sim$11\sig, $\sim$10\sig, and $\sim$6\sig\  
in the 1-3, 3-10, and 10-30~MeV bands is the quasar 3C~273 found to be on average the second most significant point source. Its fluxes in these bands are between 10\% and 15\% of the Crab flux. 
Several other extragalactic (e.g. 3C~279, PKS~0528+134, Cen~A) and 
galactic (e.g. Cyg~X-1, PSR 1509-58, a known but unidentified source at l;b:~18\deg;0\deg) sources are visible as well.
In addition there are indications for previously unknown source features
like at l;b:~75\deg;+65\deg\ in the 1-3~MeV map and at l;b:~85\deg;-65\deg\ 
in the 3-10~MeV map for example. Such spots are promising candidates for 
further dedicated analyses. 
This time-averaged approach suppresses sources which flare up
only in short time periods. Therefore the maps show fewer sources than
are listed in the COMPTEL source catalog (see \cite{Schonfelder99}). 

\begin{table}[t!]
\caption{Detection significances (\sig) in 3 different energy bands 
of some known COMPTEL AGN sources (plus Crab) as derived by this all-sky analysis for the sum of data of the first 6.5 years of the COMPTEL mission.
}
\label{table1}
\begin{tabular}{lcccc}
\gray\ Source & 1-3~MeV & 3-10~MeV & 10-30~MeV & Source Type \\
\tableline
Crab         & $\sim$110 & $\sim$76 & $\sim$44 & pulsar + nebula \\
3C~273       & 11.5      & 10.8     & 6.2      & FSRQ\tablenote{Flat-Spectrum Radio Quasar}\\
3C~279       &  5.3      &  5.7     & 4.9      & FSRQ \\
PKS~0528+134 &  ---\tablenote{Unknown due to uncertainties in removing the strong Crab signal}    &  3.0     & 6.0      & FSRQ \\
PKS~1622-297 &  $<$3     &  $<$3    & 4.1    & FSRQ  \\
3C~454.3     &  $<$3     &  5.3     & $<$3   & FSRQ  \\
Cen~A        &  5.7      &  5.3     & 3.0    & Radio Galaxy  \\
\end{tabular}\end{table}

For all bright and significant COMPTEL sources we have derived fluxes 
in our 4 standard energy bands for the different time periods mentioned 
above, and have combined them to MeV light curves and spectra. 
Some results for 3C~273 are shown as an example in Figure~\ref{fig2}.
In the 1-10~MeV energy band 3C~273 is detected in each CGRO Phase/Cycle i.e. in time periods of typically 1~year. The flux turns out to be rather 
stable and varies only within a factor of $\sim$2 in the 1-3~MeV and within 
a factor of $\sim$4 in the 3-10~MeV energy band. The spectra show the same 
trend. Whereas the flux below 3~MeV turns out to be same for both halves, 
there is an indication that at the upper COMPTEL energies ($>$3~MeV) 
the source was brighter during the second half. All three spectra 
clearly show the spectral turnover occuring at MeV-energies. 
However, we emphasize that for final conclusions a dedicated source
analysis has to be carried out.

\section*{Summary}
We have applied the maximum-likelihood method to COMPTEL all-sky data 
of different time periods. By simultaneously fitting models for the 
different diffuse emission components, this analysis method provided quantitative all-sky results -- significances and fluxes -- on point sources. 
An all-sky summary on their time-averaged fluxes 
and significances is thereby provided. After the Crab -- pulsar plus nebula -- 
the quasar 3C~273 was found to be the most significant COMPTEL MeV-source, having time-averaged fluxes of the order of 10\% to 15\% of the Crab. 
Additional evidence for previously unknown source features has been found as well.  

\vspace{0.4cm}\noindent 
{\small ACKNOWLEDGMENTS: The COMPTEL project is supported by the German government through DARA grant 50 QV 9096 8, by NASA under contract NAS5-26645, and by the Netherlands Organisation for Scientific Research (NWO).}

\end{document}